# Fuzzy Logic Based Multi User Adaptive Test System

*Atif Ali Khan[1], Oumair Naseer[2]

[1]*School of Engineering, University of Warwick, Coventry UK*
[2]*Department of Computer Science, University of Warwick, UK*
Email: [1]Atif.Khan@warwick.ac.uk , [2]O.naseer@wariwck.ac.uk

*Abstract.* The present proliferation of e-learning has been actively underway for the last 10 years. Current research in Adaptive Testing System focuses on the development of psychometric models with items selection strategies applicable to adaptive testing processes. The key aspect of proposed Adaptive Testing System is to develop an increasingly sophisticated latent trait model which can assist users in developing and enhancing their skills. Computerized Adaptive Test (CAT) System requires a lot of investment in time and effort to develop analyze and administrate an adaptive test. In this paper a fuzzy logic based Multi User Adaptive Test System (MUATS) is developed. Which is a Short Messaging Service (SMS) based System, currently integrated with GSM network based on the new psychometric model in education assessment. MUATS is not only a platform independent Adaptive Test System but also it eases the evaluation effort for adaptive test process. It further uses fuzzy logic to pick the most appropriate question from the pool of database for a specific user to be asked which makes the overall system an intelligent one.

* *Corresponding address:*
Atif Ali Khan,
*School of Engineering, University of Warwick,*
*Coventry, UK*
Email: *Atif.Khan@warwick.ac.uk*

## 1. Introduction

The aim of this project is to evaluate the Multi User Adaptive Test System (MUAST). The purpose of this tool is to facilitate the learning of school-going kids aged (7 – 15 years) or adults who have not been through formal education and lack basic mathematical skills with an option to extend it to different subjects. The system is kept open source thus making it available for everyone to use. Multi User Adaptive Test System is an adaptive game [1] through short messaging service (SMS) for educating learners who lack analytical skills [4]. Learning aspect is the most difficult part of the system. To make this system more enjoyable from the learning point of view, a lot of provisions are added to the system. This computer adaptive game is built through SMS for educating learners who lack mathematical skills [3]. Educating learners through SMS is a challenging task. Due to sequential nature of the medium, the method of teaching is of question answer format. Initially, a basic question is sent to the user. If the user answers it correctly then he/she will be forwarded similar questions to verify that the concept is correct. After a series of correct answers the learner/user is progressed to the next or more advanced level of tasks. For example moving to a higher level could mean moving from single digit addition to double digit addition. The most challenging aspect is to identify weak or missing concepts through the series of errors that the learner makes and then to help him/her grasp these very concepts from other suggested queries. When a wrong answer is received, the cognitive association of error is used with the model fault in the mind of learner. If a learner makes errors which points to a type of modeling fault then the learner is provided with some aids to correct the model [10].

The learner is re-tested on this concept and if the learner answers it correctly then the learner is moved to the next level. Moreover the system is made intelligent using fuzzy logic, so that it can decide on the difficulty level for specific user itself. The selection or switching between the difficulty levels is incorporated using the if-then rules which is the basis for fuzzy logic. Matlab fuzzy logic toolbox is used to design and implement control strategy for question selection where different inputs are fed to the controller, which decides the output using if-then fuzzy rules.

## 2. System Architecture

The system architecture is shown in figure 1 below. It consists of the Multiple Choice Questions (MCQs) database stacked on the basis of the difficulty level of the questions. A GSM based mobile phone having SMS service and a fuzzy logic based controller to make decision on the selection of difficulty level of the Multiple Choice Question (MCQ) which needs to be feed to the user/player [2]. The decision parameters for fuzzy based controller are age, education level and previous standing of the user/player. Another database [12] is maintained to keep up to date record of the user/player current standing which will help in calculating the learning curve for a specific user over a certain period of time [9].

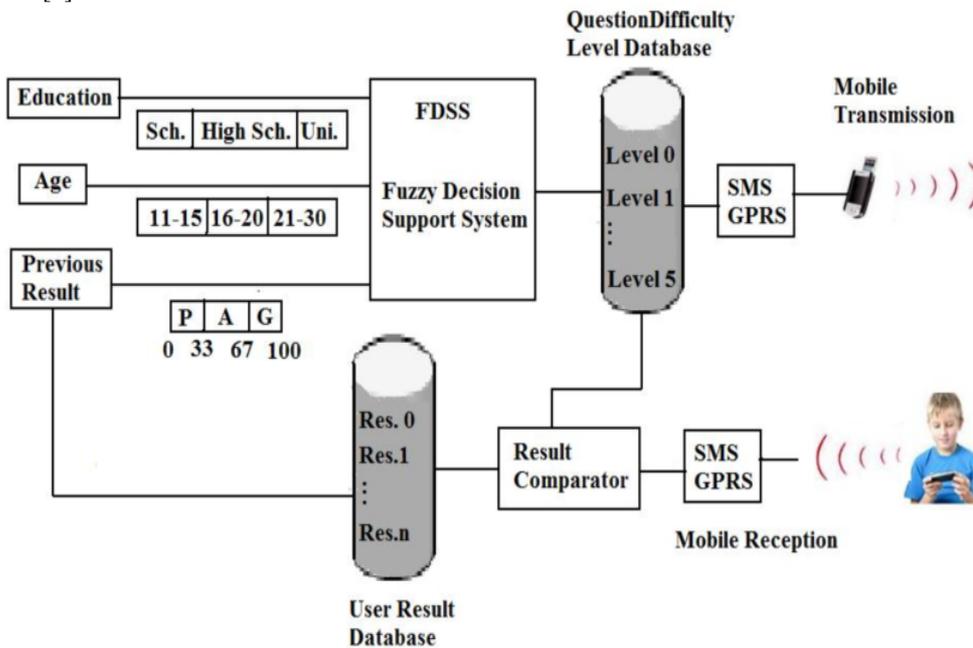

Figure 1. System Architecture of Fuzzy Logic based MUATS

The database of overall system consists of eight tables which are discussed below:

### 2.1 Active-Players

This table keeps the record of players which are active at any given instant; normally it is loaded in memory for faster access.

### 2.2 Game History

This table keeps full history of every player for the purpose of getting statistics about that player whenever necessary. The history saved about players includes the question asked, the time it took to be answered and whether the answer was correct or not. The software adapts new levels of a game based on the statistics in this table.

### 2.3 Message Buffer

This table serves as a buffer between the program and the output device. It is implemented to cater the speed difference between the two. For now, the output device is a modem thus its speed is considerably slower than that of a PC therefore the processed message responses are stored in this table and from there, the modem picks them one at a time and throws them onto the GSM network. This table has only two fields which keep the phone number (as the message recipient address) and the message text body.

### 2.4 Player

This table records the total players ever registered to play the SMS adaptive games. It has four columns:
a. Phone-No: The phone number from which the player was registered.
b. Player-ID: The Player's name supplied by the player at the time of registration.
c. Reg.-Date: The date when a player first registered to the system.
d. Last-Game-Played: A date field to keep the date when the player last played a game.

### 2.5 Player-Level

This table keeps the record of all players and their performance (total questions asked and the number of correct questions for each level) at all levels of all games ever played. These stats allow making an adaptation decision through fuzzy logic system.

### 2.6 Questions

This table is to store all the questions, their levels, answers and helps. This table is brought into memory at startup because it will be accessed frequently later on.

### 2.7 Top Level Statistics

This table makes housekeeping easy by storing available levels of a game /topic and number of questions present on that level. A question is chosen if it is present and is quickly checked by looking in this table. A question will be chosen only if it exists and using this table one can know exactly how many levels of a game are there with the total number of questions on that level.

### 2.8 Topics

This table stores the numerous topics offered by the system to play games. In SMS message on start of a game (after successful login) these topics are sent to the player to make a choice what he/she might want to play/learn. This table is kept in memory for faster access.

## 3. Methodology

The SMS user of any mobile network can register and play with the proposed system. A fuzzy logic system is implemented which allows multiple players per phone number [5]. This is done because the audience is children who cannot have a phone of their own due to regulations of Telecom Authority which require having a National Identity Card number to own a mobile phone sim. Therefore any number of children/user can play from the mobile phones of their parent which will be used for all the siblings of a family. The system allows registering up to 10 players per phone number. An example illustration of this system is shown in figure 2 below.

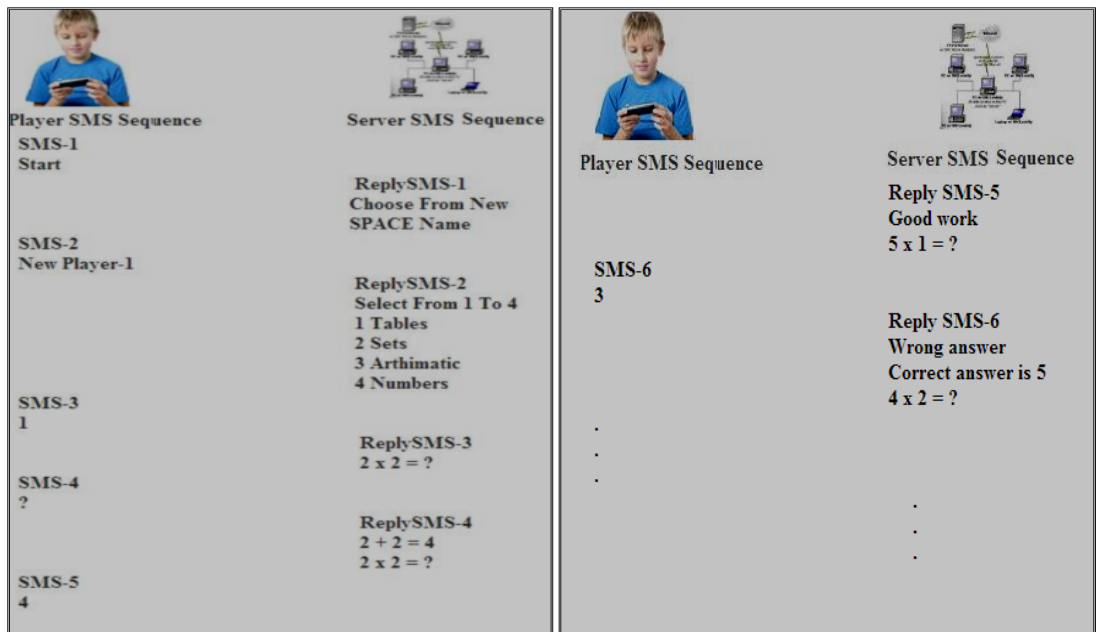

Figure 2. Example illustration of the game

All incoming messages are converted to uppercase by default and remove any preceding and proceeding spaces therefore the text we look for in a message is not case sensitive and is free of its position inside the message. The Global System for Mobile Communications (GSM) network by default gives The Phone number, Date & time, and the message text as SMS message to the system. Only the phone number and the message text to process and respond are picked.

To start a fresh game, the user needs to send an SMS which has "START" as the message body text; this will remove any previous players if active from that phone number and the system starts identifying the new incoming user. In response to the "START" command, the program will send a list of registered players from that phone number [13]. This response also includes the option to add a new player until the number of registered players for this phone number becomes 10. Only the first timer will get this option. To add a new player if allowed, the user needs to respond with a message containing "NEW [space] Name" as the message body text, where "[space]" is the space character (ASCII 27) and "Name" is name the user wants to use in future. This is how a new user registers with the system. The existing player needs to send his/her name spelled exactly as displayed in the user list message. Any other response will send the user list once again. This is done until a successful login. The valid username provided becomes an active player and is thus added to the active players table in the memory.

Now a message is sent containing a list of available topics to choose a game. From here on, the user does not need to type a language for a response and therefore the language barrier and typographic errors problem is removed. All the responses are expected to be numeric. Thus the list of topic which is sent contains topics and their ID numbers. The user response should be the ID number which he/she wants to play. Any invalid message (other than the keywords "START","NEW" and "EXIT") will cause the system to keep sending the topic list message again. Upon successful selection of a topic the system accesses the history table and computes a suitable level for the player to begin sending questions. First time, the players of a topic are considered beginners and there history is initialized to be on level one. A question is randomly selected from the computed level and sent to the user as an SMS message. Now the game has begun and system is waiting for an answer from the user.

All the questions are set to multiple choices type to ease the burden of typing on the player. The next message from this number is supposed to be the answer of the asked question so system keeps this question, its correct answer and its help in the active players table [15]. And on reception of next message from this number system checks the answer if the answer is "#" then a response is sent which contains the help to the current question otherwise if the response is a correct answer then 1 is added to the correct-answers counter for this player otherwise correct-answers counter is not incremented, the total-asked counter according to the level is also incremented in active players table and player history table.

If the total asked number exceeds the threshold of adaptation (currently 10) system adds the total asked and total correct numbers from active players table into the player history table. Then it calculates the level which should be used next based on the complete history of the player in that game and on that level. Any key word ("START","NEW" or "EXIT") or a time out causes the system to back up the current user on the phone number from the active players table into the player's history table [11]. On exit, due to the key word or a time out, the player's data of current game is added to the player history table and the player's records in the active players table are deleted. Any further message from this phone number will cause a response that the player needs to start a new game by sending a key word "START". Time out is implemented as a timer which on regular intervals checks the active players table for players which have not sent any messages during a preset time period [8].

## 4. Fuzzy Controller/ Decision Support System
### 4.1 Control Group Selection Criteria

A control group is set up where a group of few adults are provided with the questions. They are evaluated before the start of experiment to validate their level of expertise in subject matter. They are then asked to play the game. The validation will answer the following research question:
1. Is the learner able to answer the questions that he/she already knows?
2. Is the learner able to benefit from the aids/help provided at time of failure?

The first question is to validate that the questions are understandable for the learners. If the learner is not able to answer the questions then system will re-evaluate questions. The second research question validates educational goals that aid messages are able to correct an error of a learner. When a wrong answer is received, system will use the cognitive association of error with the model fault in the mind of learner. [6]

### 4.2 Fuzzy Controller Design

Fuzzy Logic works very well for the difficulty level selection of the questions according to the competence of the user/player [7]. The design of the fuzzy controller is shown in figure 3 below. The output of the fuzzy controller is "Question Difficulty Level" which is decided on the basis of three inputs Education Level, Age and Previous Standing of the user/player [14].

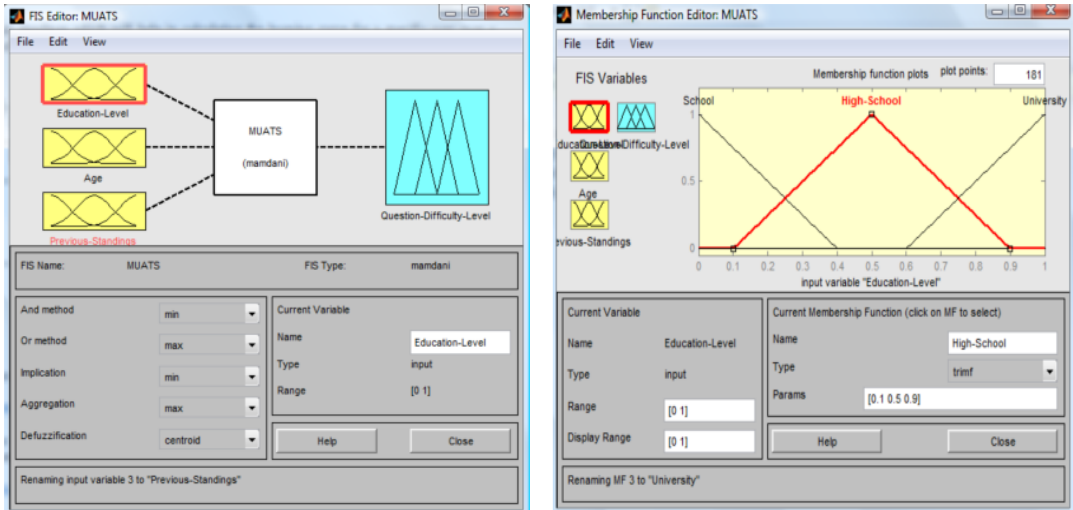

Figure 3. Mamdami Type Fuzzy System for MUATS

Inout-Output memberships functions are given below in figure 4 below, which are kept simple yet meaningful. Education level has three membership functions (School, High School and University). Age is divide into three cluster as (Child, Teen and Mid Age) having a range between 10 to 30 years. The third input Previous Standing also has three membership functions as (Poor, Average and Good). The output has six membership functions as (Level 0, Level 1, Level 2, Level 3, Level 4, Level 5) which is to estimate the appropriate difficulty level of the question for a user/player.

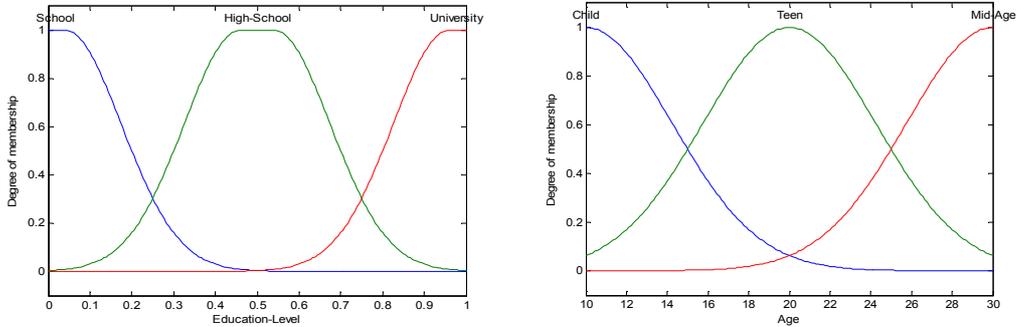

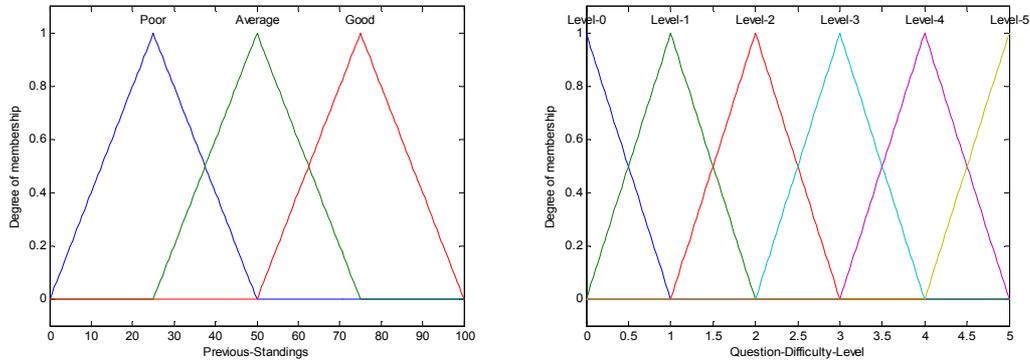

Figure 4. Input Output Membership Functions for Fuzzy MUATS

There are total 27 rules used for the fuzzy decision support system. The rules are kept simple and meaningfull. The rule base is shown in figure 5 below:

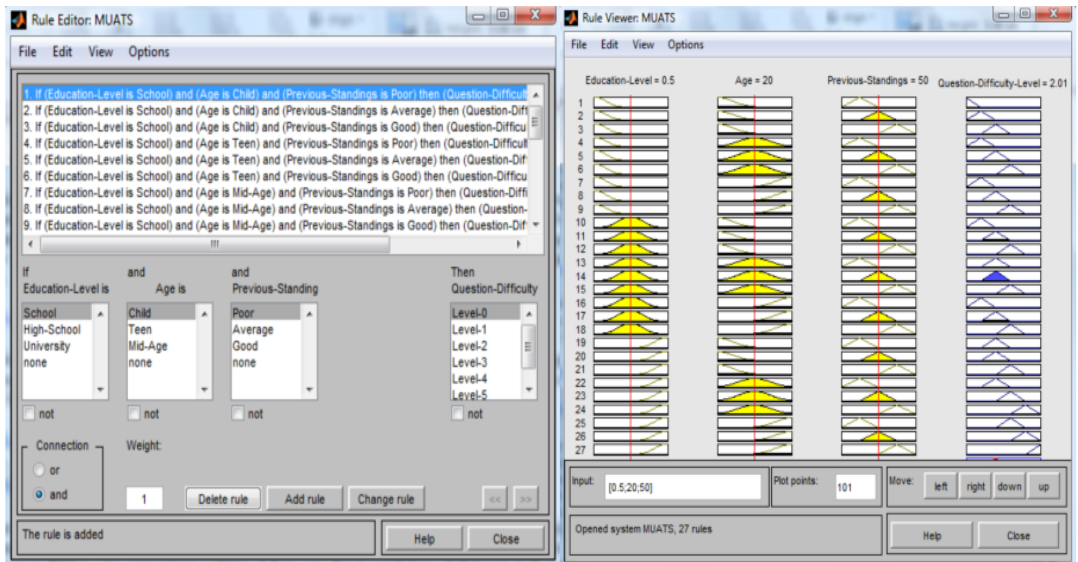

Figure 5. Rules Base for Fuzzy MUATS

The surface view of the output vs input is shown below in figure 6.

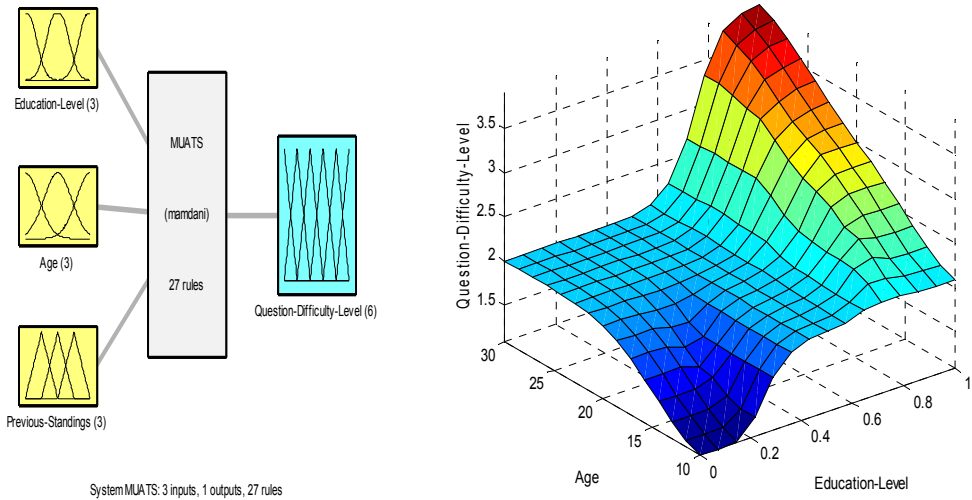

Figure 6. System Diagram and Surface View of the Fuzzy MUATS

## 5. Experiments/Evaluation

As discussed in Educational Philosophy, learning comes to play when an error is identified in a learners approach and thus through suggestive messages; the error in learners approach is corrected. To evaluate the system a series of experiments are conducted on learners who lack the skills that proposed system is trying to teach.

### 5.1 EXPERIMENT-1
#### 5.1.1 Control Group

For this experiment 4 individuals were selected from the LUMS PEPSI DINING CENTRE. The background of the control group was as follows:

*Table 1.* Control Group Description

| Name | Age | Latest Qualification | Employment History | Interest | Current Status | Future Goal |
|---|---|---|---|---|---|---|
| Mr. ATIF | 32 | 8 | LUMS, PDC | learning | Cook | cashier |
| Mr. RAYAZ | 35 | 5 | LUMS, PDC | games | Cook | cook |
| Mr. MUBEEN | 29 | 7 | LUMS, PDC | math | Cook | cashier |

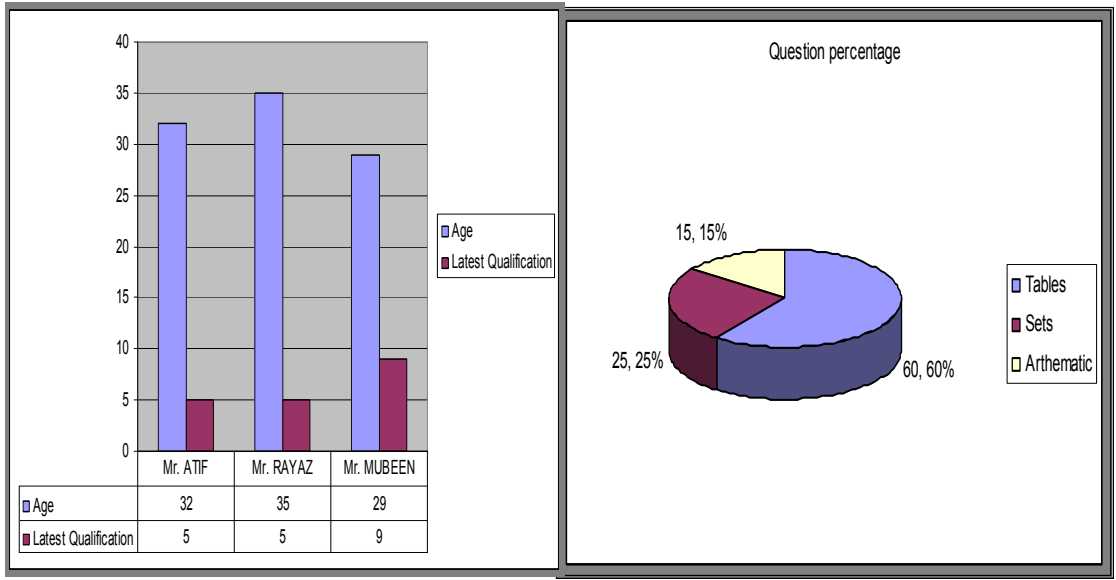

Figure 7. Relation between user's age and qualification

### 5.1.2 Statistics before and after the Experiment

At the start of the experiment a 30 minutes lecture was given to the control group. After that a 20 minutes demonstration was given on the game user guide. When the control group feels comfortable with the game, the actual experiment was carried out. The statistics before and after the experiment are shown in figure 8 below.

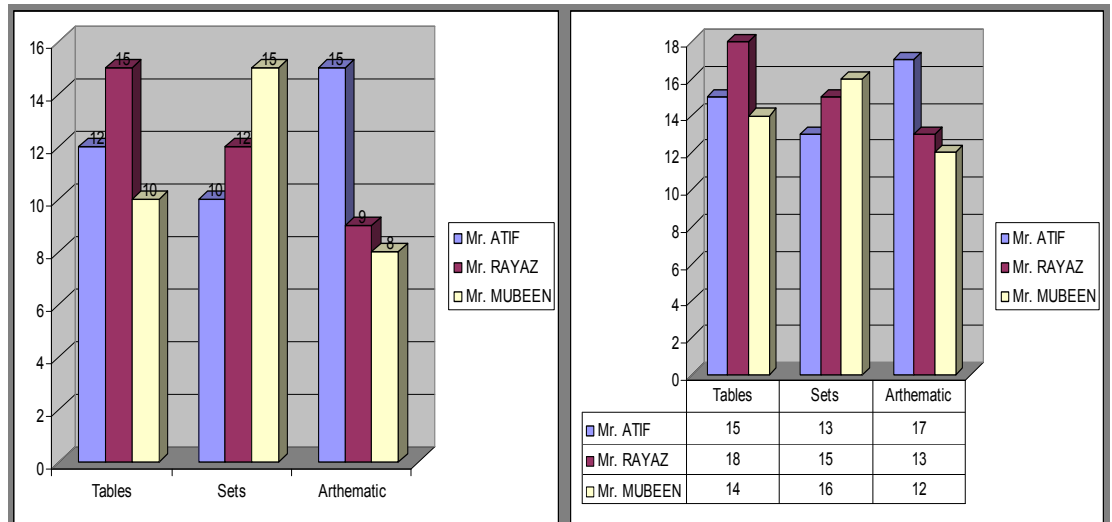

Figure 8. Statistics before and after the experiment

### 5.1.3 Results
Following graphs (figure 9) shows the improvement of each user/player after the experiment.

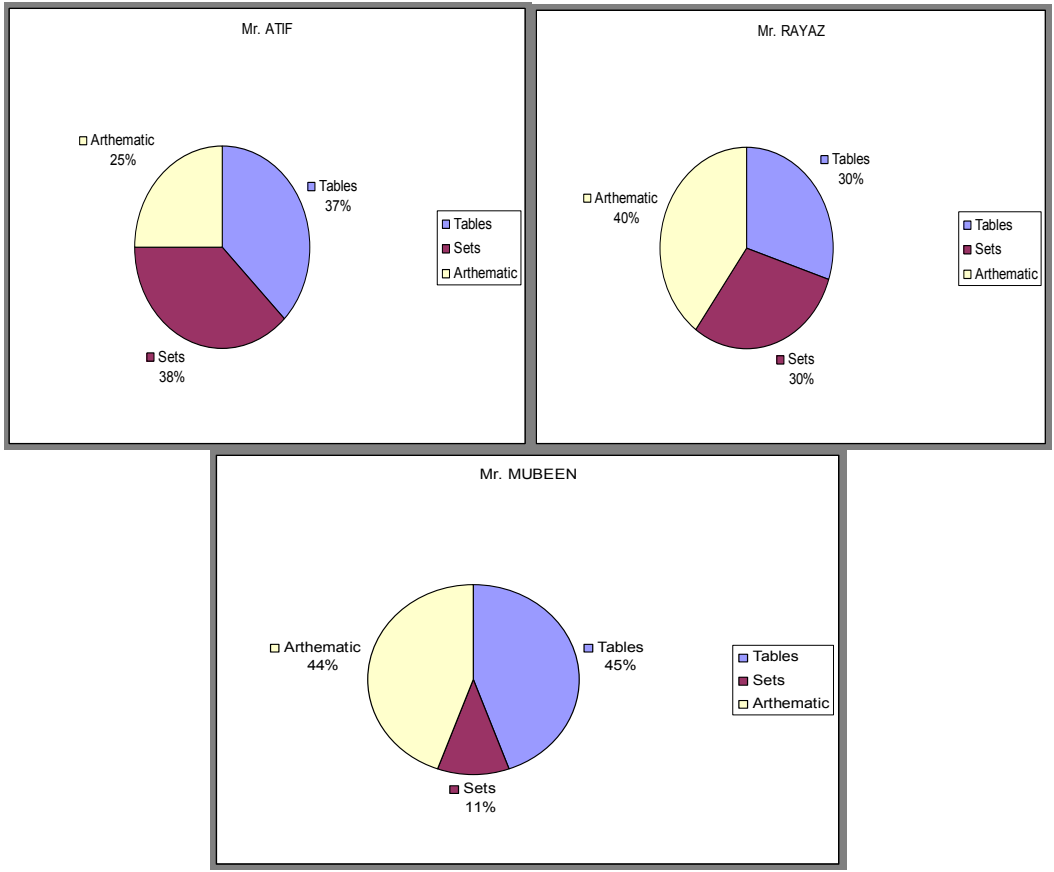

Figure 9. Improvement shown by each user/player.

### 5.2 EXPERIMENT-2
#### 5.2.1 Control Group

For this experiment 4 individuals were selected from the LUMS PEPSI DINING CENTRE. The background of the control group was as follows in table 2:

*Table 2.* Control Group description

| Name | Age | Latest Qualification | Employment History | Interest | Current Status | Future Goal |
|---|---|---|---|---|---|---|
| Mr. KHURRAM | 30 | 8 | LUMS, PDC | learning | Cook | cashier |
| Mr. RAYAZ | 35 | 5 | LUMS, PDC | games | Cook | cook |
| Mr. MUBEEN | 29 | 9 | LUMS, PDC | math | Cook | cashier |
| MR. SADIQ | 25 | 8 | LUMS, PDC | math | Cook | cashier |

Before the start of the experiment a couple of questions were asked to each user/player of the control group.

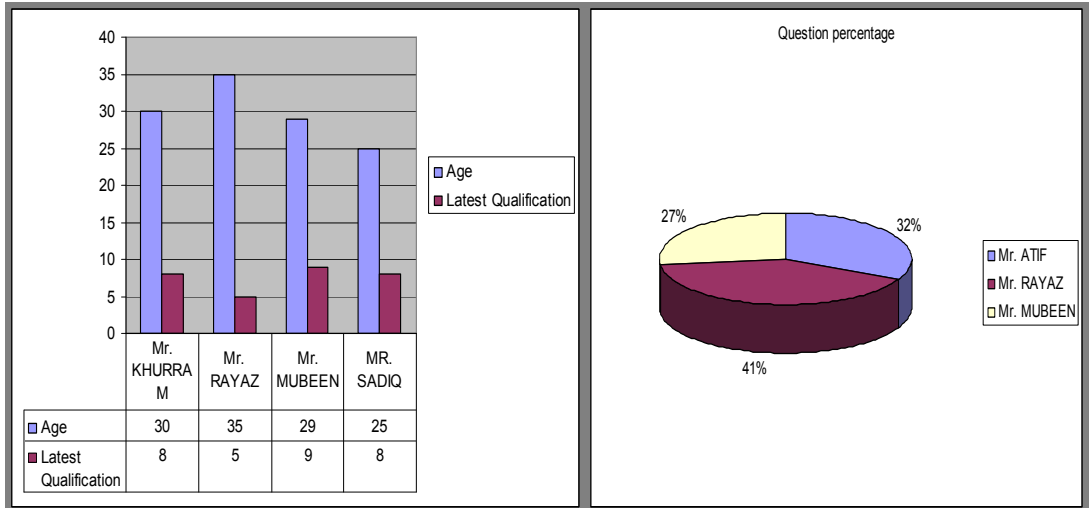

Figure 10. Relation between age and qualification

### 5.2.2 Statistics before and after the Experiment

In this experiment only the game playing guide line were demonstrated to the control group. The purpose of this experiment was the language understanding. Also a set of multiple choice question were given to test their skills.

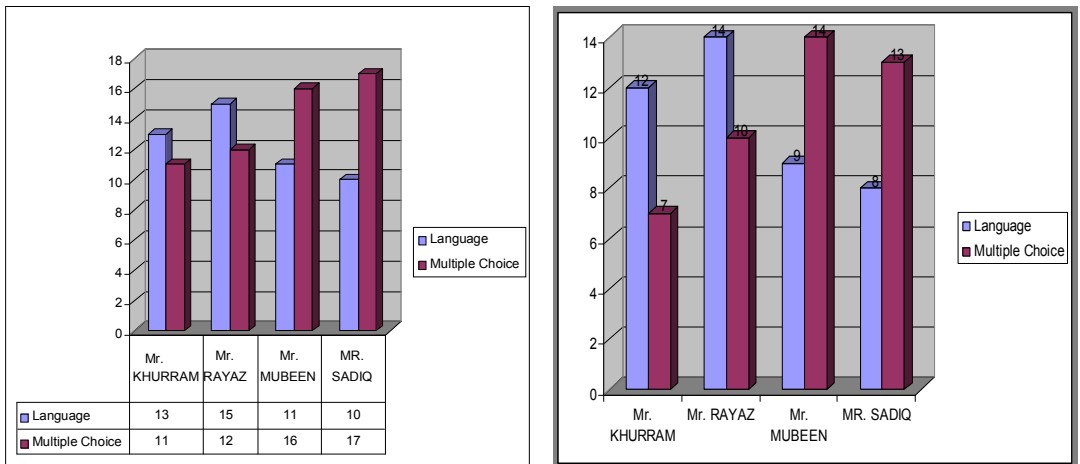

Figure 11. Statistics before and after the start of experiment

### 5.2.3 Results:

Following graphs (figure 12 and 13) shows the improvement of each user/player after the experiment.

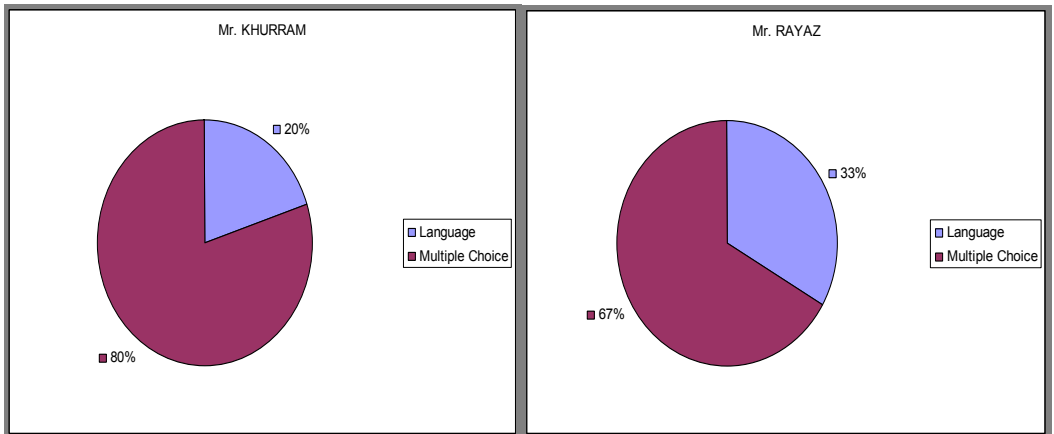

Figure 12. Improvement shown by each user/player

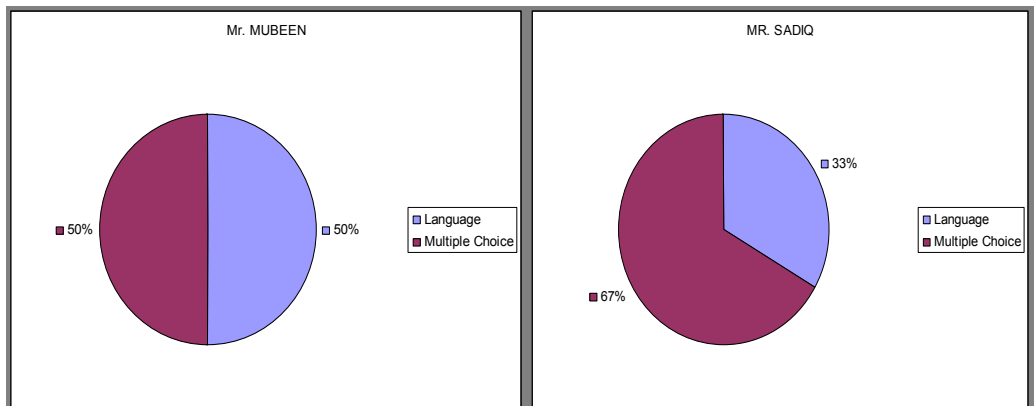

Figure 13. Improvement shown by each user/player

## 6. Results and Conclusion

Based on the age, education level, pervious knowledge and intellectual ability of the control group, MUATS has shown a significant improvement. Experiments reveal that the MUATS is very successful in identifying the misconception or weaknesses among control group of any background. During the learning process, a gradual increase in difficulty levels of the problems and the repetition of same problems helps control group to grasp the main concept from the suggested queries. CAT system requires a lot of investment, effort and time whereas the fuzzy based Multi User Adaptive Testing System proves its efficiency from the time, evaluation and administrative perspective. MUATS suggests a paradigm shift from CAT system to SMS based Adaptive testing system. MUATS is a platform independent Adaptive test system which provides integration to any mobile network. In future, integration of CAT with MUATS will explore the new horizons for researches both from the Psychometric models in education assessment and from implementation of adaptive testing systems using state of the art technologies. Though, the most challenging aspect is to identify weak or missing concepts through the series of errors that the learner makes and then to help him/her grasp these very concepts from some other suggested methodologies.

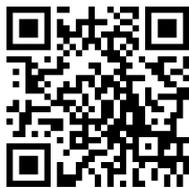

Free download this article
and more information